# A Text Classification Application: Poet Detection from Poetry

DURMUŞ ÖZKAN ŞAHİN[1], OĞUZ EMRE KURAL[1], ERDAL KILIÇ[1] and ARMAĞAN KARABİNA[1]

[1] Ondokuz Mayıs University, Samsun/Turkey,
{durmus.sahin, oguz.kural, erdal.kilic, armagan.karabina}@bil.omu.edu.tr

*Abstract* – With the widespread use of the internet, the size of the text data increases day by day. Poems can be given as an example of the growing text. In this study, we aim to classify poetry according to poet. Firstly, data set consisting of three different poets' poetry written in English have been constructed. Then, text categorization techniques are implemented on it. Chi-Square (CHI) technique are used for feature selection. In addition, five different classification algorithms are tried. These algorithms are Sequential minimal optimization (SMO), Naïve Bayes (NB), C4.5 decision tree, Random Forest (RF) and k-nearest neighbors (KNN). Although each classifier showed very different results, over the 70% classification success rate was taken by SMO technique.

## I. Introduction

WITH the widespread use of the internet, text data have augmented day by day. Therefore, produced and recorded in the type of text data have to be operated by computers. For that reason obtaining and using meaningful information from these data will facilitate the work of people. Text mining is multi discipline; it makes use of such as data mining, artificial intelligent, natural language processing and information retrieval. Text classification is one of the most widely studied areas in text mining. It is a supervised learning problem. To solve this problem, firstly, data separated into two parts called train and test data. In the next step, classification algorithms applied training data. In the last step classifier models predict label of text by means of learning model. There are many text-mining studies in the literature.

Spam mail detection: Email is the simplest and cheapest communication tools in daily life. Most of the received emails are the text. These mails are two parts, named spam and non-spam. Some advertisement companies or bad people prefer to send spam mails to users. Spam mails give rise to waste of time and slow down daily work. Therefore, spam mails have to be found and not delivered to the users. Spam mail detection is binary text classification application [1, 2].

Web page classification: Web pages occur in HyperText Markup Language (HTML) files. There are some information based on text in these files. If text-mining techniques apply to these files, web pages may classify by the computers [3].

Author detection: Text-mining methods apply to the books or column so that author of texts can be recognized [4, 5].

There are very different studies apart from the above popular topics. These studies are gender of text's authors [6], automatic text summarization [7], automatic question answering systems [8], sentiment analysis on text [9, 10], music genre classification on lyrics [11, 12], television-rating prediction with social media [13] and automatic news article classification [14 – 16].

In this study, we aim to classify poetry according to poet. The first part of the study is to generate corpus. Thus, poetry of three different poets are taken from web page [17]. Poets are Adryan Rotica, Lamar Cole and Richard Allen Beevor. Generating data set is balanced. After corpus have generated, text classification techniques were applied to the corpus.

## II. Material and Method

Although computers can read text files, classifiers cannot understand them. In addition, texts are named unstructured data. For this reason, text files have to convert into the numerical data so that they can be structured data anymore. In Figure-1 flowchart shows general text classification applications' steps.

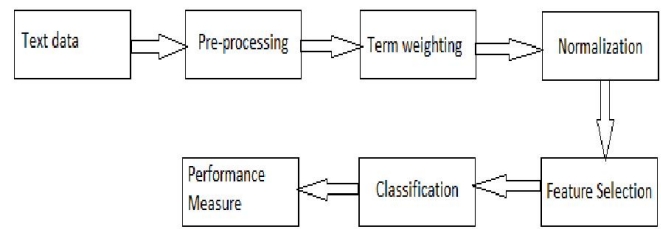

Figure 1: Text classification steps

Text files transform into structured data via pre-processing and term weighting steps. Each of steps are detailed sub sections.

### A. Pre-processing

The words in the documents are not always in the same form. These words should be converted to the same style because they have the same importance. If words is not transformed into same style, the numbers of words are different so that may lead to misleading results. In pre-processing step, all the words converted uppercase into lowercase, punctuation marks and digits deleted, tokenization done according to white space character, found stemming type all the words and stop words deleted. We prefer Porter Stemmer to find stem of words [18].

*B. Term weighting*

Any term is very few in some documents but it can be seen very much in some documents. Therefore, the term weighting must reveal this distinction. Term Frequency – Inverse Document Frequency (TF.IDF) was used as term weighting method that is frequently used in text classification. TF is the number of occurrences in a document. IDF is

$$IDF = \log\left(\frac{N}{a+c}\right) \quad (1)$$

given Equation 1. In Equation 1, $N$ is the number of documents in the whole collection, $a$ is the number of documents in the positive category that contain this term, $c$ is the number of documents in the negative category that contain this term. The weight of the term is calculated by multiplying TF by IDF.

*C. Feature selection*

The operation of classification algorithms takes a long time. At the same time, the operation of these algorithms is directly related to the size of the data. Bag of words model is preferred in traditional text categorization applications. Document vectors are created using all the words in the training set. For this reason, documents are represented in terms of tens of thousands. Majority of these terms are not considered to affect important information because they do not affect the classification success positively. Instead of using all the terms in the bag of words, little word vector is preferred. Little vector is the best subset of bag of words. By means of feature selection, memory wastage can be prevented by reducing vector size. Runtime of the text classification process reduce.

   Feature selection methods are generally divided into three main groups [19, 20]. These are filtering, wrapper and embedded methods. In feature selection step, one of the filtering method CHI is used. CHI is

$$CHI(t_j, c_i) = N \frac{(ad - bc)^2}{(a+c)(b+d)(a+b)(c+d)} \quad (2)$$

given Equation 2. In Equation 2, the CHI score of the $t_j$ term in class $c_i$ is calculated. $N$, $a$ and $c$ is emphasized in Equation 1. $b$ is the number of documents in the positive category that do not contain this term and $d$ is the number of documents in the negative category that do not contain this term. The CHI scores of all terms are calculated. Subsequently these values of CHI sort big to small. The greatest value is the most relevant with related category. If CHI value of any term is 0, it is not relevant with related category.

*D. Classification*

WEKA is data mining tool that is used in the classification stage [21]. There are a lot of data mining and machine learning algorithms in WEKA tools. SMO based Support Vector Machine (SVM), NB based probability model, C4.5 and RF based decision tree and KNN are used in this study.

*E. Data set*

Table-1 shows poems and their poetry distribution. Data set divided approximately 60% train and 40% test. Because of the number of poetry is close this data set is balanced.

Table 1: Poems-poetry distribution

| Poems | Train | Test |
|---|---|---|
| Adryan Rotica | 284 | 189 |
| Lamar Cole | 241 | 162 |
| Richard Allen Beevor | 227 | 152 |

*F. Performance Measure*

With the performance measurement, the accuracy of belonging to the relevant class is evaluated. A sample that is actually labeled positively in the dataset, if it is classified as positive in the classification result, it is named True-Positive (TP). A sample that is actually labeled negative in the dataset, if it is classified as negative in the classification result, it is named True-Negative (TN). A sample that is actually labeled negative in the dataset, if it is classified as positive in the classification result, it is named False-Positive (FP). A sample that is actually labeled positive in the dataset, if it is classified as negative in the classification result, it is named False-Negative (FP). These states are shown on the Table-2 as confusion matrix.

Table 2: Confusion matrix

| Real / Predict | C1 | C2 |
|---|---|---|
| C1 | TP | FP |
| C2 | FN | TN |

The most used performance measure in text classification is F-score. F-score is the harmonic mean of precision and recall values. Precision ($\pi$) and recall ($\rho$) are given Equation 3 and Equation 4 respectively.

$$\pi = \frac{TP}{TP + FP} \quad (3)$$

$$\rho = \frac{TP}{TP + FN} \quad (4)$$

If harmonic mean of $\pi$ and $\rho$ values are calculated,

$$F = \frac{2\pi\rho}{\pi + \rho} \quad (5)$$

F-score is taken place as Equation 5. To evaluate classification success F-score is preferred.

## III. RESULTS

   The performance results of 5 different classifiers with different working models are given in Figure 2.

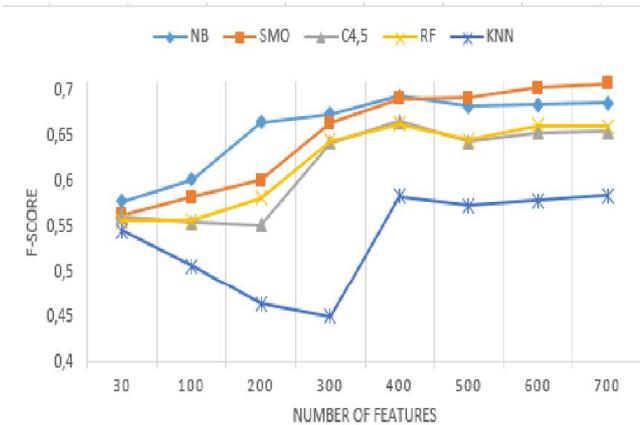

Figure 2: Success of classification algorithms

When 30 features used, the best classifier is NB. However, the worst classifier is KNN. As long as number of features increase, the success rate of all classifiers excluding KNN go up. The best classification rate is 0,7067 F-score that occurs with SMO in 700 features. Success of RF and C4.5 is similar but they don't exceed KNN and SMO.

As a result, English poetry were classified according to poets. If we use larger data set, more effective text classification results can be obtained. Instead of traditional text classification methods, semantic text classification method may be preferred. Because, poets use very special sentences while writing poetry. As a future work, we aim to generate different poetry data set to classify genre of poetry with text classification.

## References

[1] A. M. K. Chae, A. Alsadoon, P. W. C. Prasad and S. Sreedharan, "Spam filtering email classification (SFECM) using gain and graph mining algorithm" 2017 2nd International Conference on Anti-Cyber Crimes (ICACC), Abha, pp. 217-222, 2017.

[2] T. Vyas, P. Prajapati and S. Gadhwal, "A survey and evaluation of supervised machine learning techniques for spam e-mail filtering" 2015 IEEE International Conference on Electrical, Computer and Communication Technologies (ICECCT), Coimbatore, pp. 1-7, 2015.

[3] S. A. Özel, "A Web page classification system based on a genetic algorithm using tagged-terms as features", Expert Systems with Applications, 38(4), pp. 3407-3415, 2011.

[4] M. Yasdi and B. Diri, "Author recognition by Abstract Feature Extraction" 2012 20th Signal Processing and Communications Applications Conference (SIU), Mugla, pp. 1-4, 2012.
doi: 10.1109/SIU.2012.6204690

[5] P. Tüfekci and E. Uzun, "Author detection by using different term weighting schemes" 2013 21st Signal Processing and Communications Applications Conference (SIU), Haspolat, pp. 1-4, 2013.
doi: 10.1109/SIU.2013.6531190

[6] S. Doğan, B. Diri, "Türkçe Dokümanlar için N-gram Tabanlı Yeni Bir Sınıflandırma: Yazar, Tür ve Cinsiyet" Türkiye Bilişim Vakfı Bilgisayar Bilimleri ve Mühendisliği Dergisi, 4(4), pp. 83-92, 2010.

[7] A. Güran, S. N. Arslan, E. Kılıç and B. Diri, "Sentence selection methods for text summarization" 2014 22nd Signal Processing and Communications Applications Conference (SIU), Trabzon, pp. 192-195, 2014.
doi: 10.1109/SIU.2014.6830198

[8] M. Amasyalı, B. Diri, "Bir Soru Cevaplama Sistemi: BayBilmiş", Türkiye Bilişim Vakfı Bilgisayar Bilimleri ve Mühendisliği Dergisi, 1, pp. 37-51, 2005.

[9] C. Parlak, B. Diri, "Farklı Veri Setleri Arasında Duygu Tanıma Çalışması" Dokuz Eylül Üniversitesi Mühendislik Fakültesi, Mühendislik Bilimleri Dergisi, 16(48), pp. 21-29, 2014.

[10] M. Meral and B. Diri, "Sentiment analysis on Twitter," 2014 22nd Signal Processing and Communications Applications Conference (SIU), Trabzon, pp. 690-693, 2014.
doi: 10.1109/SIU.2014.6830323

[11] Ö. Çoban and I. Karabey, "Music genre classification with word and document vectors" 2017 25th Signal Processing and Communications Applications Conference (SIU), Antalya, pp. 1-4, 2017.
doi: 10.1109/SIU.2017.7960145

[12] Ö. Çoban and G. T. Özyer, "Music genre classification from Turkish lyrics" 2016 24th Signal Processing and Communication Application Conference (SIU), Zonguldak, pp. 101-104, 2016.
doi: 10.1109/SIU.2016.7495686

[13] C. Akarsu and B. Diri, "Turkish TV rating prediction with Twitter" 2016 24th Signal Processing and Communication Application Conference (SIU), Zonguldak, Turkey, pp. 345-348, 2016.
doi: 10.1109/SIU.2016.7495748

[14] M. F. Amasyali and T. Yildirim, "Automatic text categorization of news articles" Proceedings of the IEEE 12th Signal Processing and Communications Applications Conference, pp. 224-226, 2004.
doi: 10.1109/SIU.2004.1338299

[15] P. Tüfekci, E. Uzun and B. Sevinç, "Text classification of web based news articles by using Turkish grammatical features" 2012 20th Signal Processing and Communications Applications Conference (SIU), Mugla, pp. 1-4, 2012.
doi: 10.1109/SIU.2012.6204565

[16] E. Kiliç, M. R. Tavus and Z. Karhan, "Classification of breaking news taken from the online news sites" 2015 23nd Signal Processing and Communications Applications Conference (SIU), Malatya, pp. 363-366, 2015.
doi: 10.1109/SIU.2015.7129834

[17] https://www.poemhunter.com/ebooks/ Last access: October 2017

[18] https://tartarus.org/martin/PorterStemmer/ Last access: October 2017

[19] Gunal, S.; "Hybrid feature selection for text classification", Turkish Journal of Electrical Engineering and Computer Sciences, vol. 20, pp. 1296-1311, 2012.

[20] Uysal, A. K.; Gunal, S., "Text classification using genetic algorithm oriented latent semantic features", Expert Systems with Applications, vol. 41, pp. 5938-5947, 2014.

[21] https://www.cs.waikato.ac.nz/ml/weka/ Last access: October 2017.